# Nanocavity-Enhanced Second-Harmonic Generation from Colossal Quantum Dots

*David Sharp,*[1] *Abhinav Kala,*[1] *Hannah Rarick,*[1] *Hao A. Nguyen,*[2] *Elise Skytte,*[2] *Brandi M. Cossairt,*[2] *Arka Majumdar*[1,3*]

1 Department of Physics, University of Washington, Seattle, WA 98195

2 Department of Chemistry, University of Washington, Seattle, WA 98195

3 Department of Electrical and Computer Engineering, University of Washington, Seattle, WA 98195

* arka@uw.edu

**Abstract**

Colloidal quantum dots (QDs) are an attractive medium for nonlinear optics and deterministic heterogeneous integration with photonic devices. Their intrinsic nonlinearities can be strengthened further by coupling QDs to low mode-volume photonic nanocavities, enabling low-power, on-chip nonlinear optics. In this paper, we demonstrated cavity-enhanced second harmonic generation via integration of colossal QDs with a silicon nitride nanobeam cavity. By pumping the cavity-QD system with an ultrafast pulsed laser, we observed a strong second harmonic generation from the cavity-coupled QD, and we estimate an enhancement factor of ~3,040. Our work, coupled with previously reported deterministic positioning of colossal QDs, can enable a scalable QD-cavity platform for low-power nonlinear optics.

**Introduction**

Nonlinear optical phenomena are crucial to integrated photonics applications such as laser wavelength conversion and amplification and show promise as a mechanism for all-optical switching.[1–3] However, most foundry-compatible materials exhibit weak intrinsic nonlinearity, limiting their potential for low-power on-chip integration. Furthermore, a material's crystal structure must break inversion symmetry to have a finite second-order susceptibility. This precludes silicon and silicon compounds from second-order nonlinear optical processes such as second-harmonic generation (SHG), which finds applications in spectroscopy, frequency conversion, and frequency combs.[4–6] The emergence of low-dimensional materials has opened new routes to low-power nonlinear optics, especially second-order, owing to relaxed phase matching conditions and intrinsic broken inversion symmetries.[7,8] Among these materials, colloidal quantum dots (QDs) are a promising class of materials for on-chip nonlinear optics due to their tunability, strong nonlinearity, and ease of integration with photonic devices via recent developments in deterministic heterogeneous integration techniques.[9–13] Additionally, the large bulk second-order nonlinearities of II-VI elements commonly used in colloidal QDs, such as CdSe and CdS, indicates II-VI colloidal QDs hold promise for SHG.[14]

Nanocavity integration is critical to reducing the power thresholds for nonlinear optics by confining light in a small volume for an extended period.[15] Amongst low-dimensional materials, previous works have reported cavity-enhanced SHG from monolayers of transition metal dichalcogenides,[16–18] while colloidal QDs remain unexplored for cavity-enhanced SHG. The most relevant reports have been limited to non-colloidal II-VI nanomaterials integrated plasmonic structures, with the highest cavity-enhancement factor ~1,000 in a CdS nanobelt coupled to an Au thin film.[19,20] Plasmonic structures are not favorable for low-power, scalable operations due to inherent absorptive losses from metal, and there are no reports of cavity-enhanced colloidal QD SHG in a dielectric nanocavity. Here, colloidal colossal CdSe/CdS QDs were integrated with a silicon nitride (SiN) nanobeam cavity, chosen for its high-quality factor, small mode volume, and compatibility with colloidal QD integration.[21] By comparing the colossal QD SHG generated from the cavity to that from QDs on an unpatterned substrate, we estimate a cavity-enhancement factor of ~ 3,040.

**Methods**

The nanobeam cavity, as depicted in Figure 1a, was designed according to previously established methods.[22] The cavity geometry consisted of a 330 nm thick and 650 nm wide SiN waveguide on buffer oxide with poly(methyl methacrylate) (PMMA) encapsulation. The cavity region was comprised of a one-dimensional array of elliptical holes with a major radius of 300 nm, a minor radius of 60 nm, and a period of 234 nm. Starting from the center, the period of the holes was quadratically tapered to 243 nm over 10 periods on both sides to gradually introduce the mirror region. An additional 20 holes of the same period were added to either side to increase the cavity mirror reflectivity. This design was verified in the Lumerical finite-difference time-domain (FDTD) solver and resulted in a theoretical quality factor, $Q$, of 40,000 and mode volume 2.9 $(\lambda/n)^3$, where $\lambda$ and $n$ are the resonant wavelength and refractive index of SiN, respectively. Figure 1b shows the simulated electric-field intensity profile of the targeted transverse-electric polarized cavity mode. This design was fabricated using 100 keV electron-beam lithography and a fluorine-based dry etch chemistry according to previous protocols.[23] The cavity is coupled to a waveguide and grating couplers to facilitate characterization.

Colossal core/shell CdSe/CdS colloidal QDs were synthesized using a previously reported method.[24] First, wurtzite CdSe QD cores with an average particle diameter of 3.5 nm were prepared following an established method, where particle growth was terminated 45 seconds after the injection of the Se solution.[25] The QDs were purified by centrifugation twice using hexane as the solvent and methyl acetate as the antisolvent, then stored in hexane. The QD cores were shelled with 80 monolayers (MLs) of CdS using the stepwise shelling method in increments of 30, 50, and 80 MLs, without purification between steps.[24] The final product, CdSe QDs shelled with 80 CdS MLs, was purified by centrifugation with hexane to remove small CdS particles and unreacted precursors and was stored in hexane. The colossal QDs' morphology was analyzed by transmission electron microscopy, as depicted in Figure 1c, which indicates a hexagonal diamond shape with average height of 72 nm. Colossal QDs were integrated with the nanobeam cavities by drop-casting a solution containing many QDs on the cavity chip. After allowing the solution to dry, the chip was spin coated with PMMA to encapsulate the cavities cladded with colossal QDs.

**Results and Discussion**

The nanobeam cavity was first characterized using a confocal microscope setup. In the experimental geometry used herein, a pump laser was shined on one grating coupler and the light scattered out of plane was collected from the cavity region, all using a 40x (NA 0.6) objective lens. Figure 2a depicts the cavity spectrum measured from the nanobeam cavity considered in the rest of this work. By fitting to a Lorentzian lineshape, the cavity has a resonant wavelength 808.2 nm and $Q \sim 4,500$. Next, the SHG response of the colossal QDs was collected by shining a Ti:sapphire ultrafast pulsed pump laser (Spectra-Physics Tsunami HP) on an unpatterned section of the chip. By comparing the pump laser spectrum, Figure 2b, and the collected colossal QD SHG spectrum, Figure 2c, it is evident that there is a strong SHG response from the colossal QD film as the SHG central wavelength, linewidth, and Gaussian profile are all consistent with SHG pumped by the Ti:sapphire laser. Here, the average pump power was 30 mW as measured in front of the objective lens.

The cavity-coupled QD SHG was characterized using the same experimental geometry as before by shining the Ti:sapphire pump laser on a grating coupler and collecting the cavity-coupled QD SHG signal emitted out of plane. Figure 3a depicts a representative cavity-coupled QD SHG spectrum. The spectrum shows both Gaussian and Lorentzian components as previously observed.[16] The Gaussian part is attributed to background uncoupled colossal QD SHG, and the Lorentzian part is attributed to the actual cavity-coupled QD SHG. Fitting to the Lorentzian component reveals the cavity mode is centered at 404.1 nm with a similar linewidth to the fundamental mode, which is consistent with cavity-coupled SHG.[17] The power series of the cavity-coupled QD SHG spectrum as a function of pump power is depicted in Figures 3b and 3c. In Figure 3c, the integrated counts are calculated by integrating only the Lorentzian part of each spectrum. The error bars were calculated according to one standard deviation error in each of the fits. The integrated counts follow a quadratic dependence on the pump power, as expected for a second-order nonlinear process.[26]

To calculate the strength of cavity enhancement of the colossal QD SHG, an estimate of the pump laser power coupled to the cavity was first calculated following a previously established method.[17] The intracavity power, $P_{cav}$, is calculated as:

$$P_{cav} = P \times t_w \times g \times f \qquad (1)$$

where $P$ is the nominal pump laser power, $t_w$ is the coupling efficiency of the waveguide to the cavity, $g$ is the grating coupler efficiency, and $f$ is the spectral overlap of the cavity mode and pump laser. From input-output relations (see Supplemental Information), $t_w = 0.35$, whereas FDTD simulations give $g = 0.11$, and the overlap integral $f = 0.087$. The cavity enhancement factor, $E$, is then given as:

$$E = \frac{I_{cav}}{P_{cav}^2} \frac{P_{ref}^2}{I_{ref}} \qquad (2)$$

where, respectively, $I_{cav}$ and $I_{ref}$ are the integrated counts for the cavity-coupled QD SHG and reference colossal QDs on an unpatterned section of the cavity chip, and $P_{cav}$ and $P_{ref}$ are the corresponding pump laser powers. By comparing the cavity-coupled QD SHG integrated counts for a nominal pump laser power of 40 mW (Figure 3c, $I_{cav}$ = 63, $P_{cav}$ = 132 µW) to that of the uncoupled colossal QD SHG (Figure 2c, $I_{ref}$ = 1070, $P_{ref}$ = 30 mW), the enhancement factor $E \sim$ 3,040. This calculation assumes maximum experimental collection efficiency of the cavity-QD SHG, so it is therefore a lower bound of the true enhancement factor.

## Conclusion

CdSe/CdS colloidal colossal QDs were integrated with a SiN nanobeam cavity and ~3,040 times cavity-enhancement of the intrinsic colossal QD SHG was demonstrated. This is, to the best of our knowledge, the first report of cavity-enhanced colloidal QD SHG from coupling to a dielectric nanocavity, which is expected to be more scalable than plasmonic devices. Furthermore, it has been demonstrated recently that colossal QDs can be deterministically integrated with photonic devices.[13] Future works could improve on this result by modifying both the QDs and the cavity design. Varying the QD core size is expected to have a trivial enhancement of the intrinsic QD SHG and other II-VI materials may offer stronger second-order susceptibilities.[27,28] Doubly-resonant cavities covering the fundamental and SHG wavelengths would further increase the cavity enhancement of the QD SHG.[29] This work is a step forward for low-power on-chip nonlinear optics via integration of novel materials.

## Author Contributions

D.S. designed and fabricated the nanobeam cavities and performed the optical measurements under the supervision of A.M. H.N. synthesized the QDs and E.S. performed the transmission electron microscopy under the supervision of B.C. D.S. performed the data analysis with assistance from A.K. and H.R. All authors gave input on writing.

## Funding

This material is based upon work supported by the U.S. National Science Foundation Science and Technology Center (STC) for Integration of Modern Optoelectronic Materials on Demand (IMOD) under Cooperative Agreement No. DMR-2019444. Sample fabrication and microscopy was conducted at the Washington Nanofabrication Facility/Molecular Analysis Facility, a National Nanotechnology Coordinated Infrastructure (NNCI) site at the University of Washington with partial support from the U.S. National Science Foundation via awards NNCI-1542101 and NNCI-2025489.

**Conflict of Interest**

The authors declare no competing financial interest.

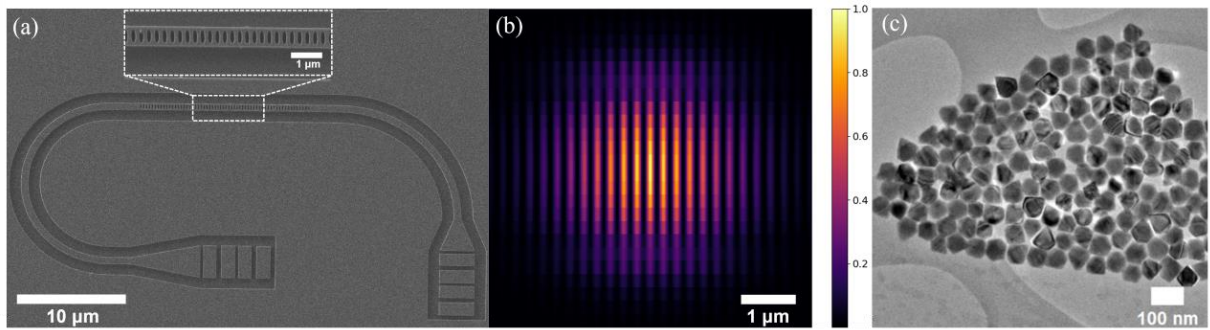

**Figure 1:** Overview of the cavity-QD system. (a) Scanning electron micrograph (SEM) of a nanobeam cavity. Inset: magnified SEM of the cavity region. (b) Finite-difference time-domain simulation of the nanobeam cavity mode electric field intensity, $|E|^2$. (c) Transmission electron micrograph of the colossal CdSe/CdS QDs used in this work.

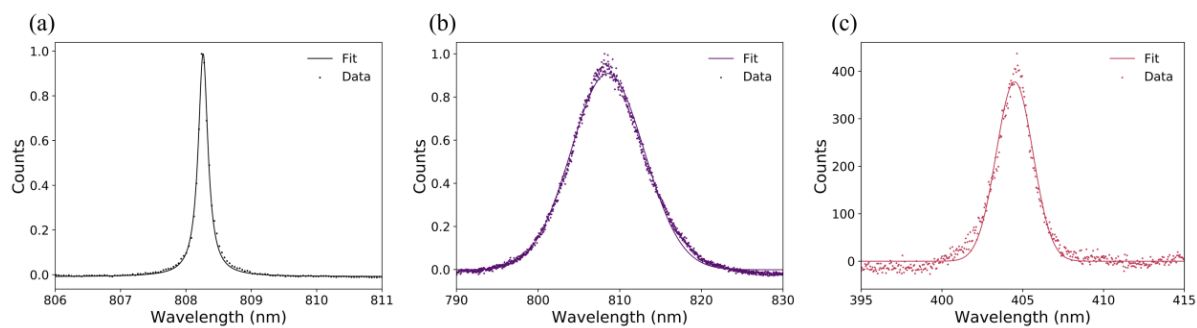

**Figure 2:** Characterization of the nanobeam cavity and colossal QDs. (a) Transmission spectrum of the nanobeam cavity as measured by shining the pump laser on a grating coupler and measuring the scattered light from the top. The fundamental mode has a resonant wavelength of 808.2 nm and linewidth 0.18 nm as determined by fitting to a Lorentzian lineshape. (b) Spectrum of the ultrafast pump laser centered at 808.7 nm with a Gaussian lineshape. (c) SHG spectrum of uncoupled QDs with 30 mW pump power centered at 404.4 nm. The central wavelength, linewidth, and Gaussian lineshape are consistent with SHG.

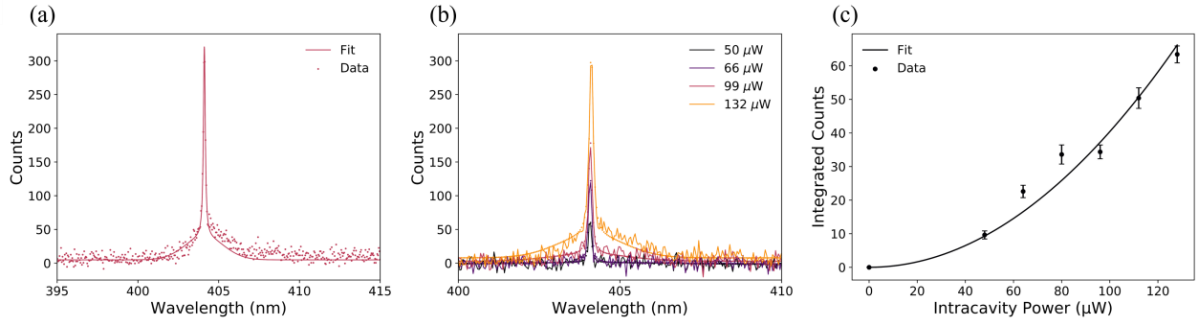

**Figure 3:** Characterization of the cavity-coupled SHG. (a) Representative QD-cavity SHG spectrum. The spectrum is fitted to the sum of Gaussian and Lorentzian lineshapes to account for the background uncoupled SHG and cavity-coupled SHG, respectively. The cavity mode is centered at 404.1 nm. The Lorentzian lineshape and central wavelength confirm this signal originates from the fundamental cavity mode. (b) Plot of the QD-cavity SHG spectrum and fit with different intracavity powers. (c) Power series of the QD-cavity SHG intensity versus intracavity power as measured by integrating the Lorentzian portion of the spectral fitting at each pump power. Error bars were obtained from one standard deviation error of each Lorentzian fit. The power series follows a quadratic trend.